\documentclass[aps,prd,floats,nofootinbib,preprintnumbers]{revtex4}

\usepackage[utf8]{inputenc}
\usepackage{mathrsfs}
\usepackage{amssymb}
\usepackage{amsmath}
\usepackage{feynmp}
\usepackage{slashed}
\usepackage{latexsym}
\usepackage{graphicx}
\usepackage{verbatim}
\usepackage{bbold}
\usepackage[normalem]{ulem}

\newcommand{\be}{\begin{equation}}
\newcommand{\ee}{\end{equation}}
\newcommand{\ba}{\begin{array}}
\newcommand{\ea}{\end{array}}
\newcommand{\bea}{\begin{eqnarray}}
\newcommand{\eea}{\end{eqnarray}}
\newcommand{\balg}{\begin{align}}
\newcommand{\ealg}{\end{align}}
\newcommand{\bit}{\begin{itemize}}
\newcommand{\eit}{\end{itemize}}
\newcommand{\trm}[1]{\textrm{#1}}
\newcommand{\mbf}[1]{\mathbf{#1}}

\newcommand{\mcl}[1]{\mathcal{#1}}
\newcommand{\mbb}[1]{\mathbb{#1}}
\newcommand{\msc}[1]{\mathscr{#1}}

\newcommand{\Mpc}{\trm{\Mpc}}
\newcommand{\yr}{\trm{\yr}}
\newcommand{\eV}{\trm{\eV}}

\newcommand{\nn}{\nonumber}

\newcommand{\tr}[1]{\trm{Tr}\left[ {#1} \right]}

\begin{document}

\preprint{NUHEP-TH/15-0X}

\title{New Chiral Fermions, a New Gauge Interaction, Dirac Neutrinos, and Dark Matter}

\author{Andr\'e de Gouv\^ea}
\affiliation{Northwestern University, Department of Physics \& Astronomy, 2145 Sheridan Road, Evanston, IL~60208, USA}

\author{Daniel Hern\'{a}ndez}
\affiliation{Northwestern University, Department of Physics \& Astronomy, 2145 Sheridan Road, Evanston, IL~60208, USA}

\begin{abstract}

  We propose that all light fermionic degrees of freedom, including the Standard Model (SM) fermions and all possible light beyond-the-standard-model fields, are chiral with respect to some spontaneously broken abelian gauge symmetry. Hypercharge, for example, plays this role for the SM fermions. We introduce a new symmetry, $U(1)_{\nu}$, for all new light fermionic states. Anomaly cancellations mandate the existence of several new fermion fields with nontrivial $U(1)_{\nu}$ charges. We develop a concrete model of this type, for which we show that (i) some fermions remain massless after $U(1)_{\nu}$ breaking -- similar to SM neutrinos -- and (ii) accidental global symmetries translate into stable massive particles -- similar to SM protons. These ingredients provide a solution to the dark matter and neutrino mass puzzles assuming one also postulates the existence of heavy degrees of freedom that act as ``mediators" between the two sectors. The neutrino mass mechanism described here leads to parametrically small Dirac neutrino masses, and the model also requires the existence of at least four Dirac sterile neutrinos. Finally, we describe a general technique to write down chiral-fermions-only models that are at least anomaly-free under a U(1) gauge symmetry.
\end{abstract}

\maketitle

\setcounter{equation}{0}
\setcounter{footnote}{0}

\section{Introduction}

All confirmed fundamental fermion fields are chiral. We can express all known matter fields as left-handed Weyl fermions -- $Q$, $u^c$, $d^c$, $L$, $e^c$ -- that transform under the Standard Model (SM) gauge group $SU(3)\times SU(2)\times U(1)$ as $(\mbf{3},\, \mbf{2})_{1/6}$, $(\bar{\mbf{3}},\, \mbf{1})_{-2/3}$, $(\bar{\mbf{3}},\, \mbf{1})_{1/3}$, $(\mbf{1},\, \mbf{2})_{-1/2}$ and $(\mbf{1},\, \mbf{1})_{1}$, respectively. There are no vector-like fermions, i.e., pairs of fields with `equal-but-opposite' charges. One consequence of this experimental fact is that, before electroweak symmetry breaking (EWSB), all fundamental fermions are massless. A corollary is that all fermion masses are proportional to the parameter that controls EWSB. In the case of the SM, this is the vacuum expectation value $v$ of the neutral component of the scalar Higgs field $H$, $(\mbf{1},\, \mbf{2})_{1/2}$. 

One can speculate why, to date, all identified fermion fields turned out to be chiral. On one hand, the existence of a single chiral fermion charged under a gauge symmetry implies the existence of several other chiral fields in order to cancel the gauge anomalies. On the other hand, vector-like fermions, if they were to exist, would have masses unrelated to EWSB and could be out of the reach of current experimental probes. The lack of concrete, unambiguous, experimental evidence leads one to conclude that, if they do exist, vector-like fermions are either very heavy or, in the event that the new fermion is not charged under the SM gauge group, very weakly coupled.

Yet, we know the SM is not complete. The SM fermion content, combined with the SM Higgs sector, leads to the prediction that neutrino masses are zero. While this is not a bad approximation -- neutrino masses are known to be tiny -- it is factually incorrect \cite{SuperK, SNO, Agashe:2014kda,deGouvea:2013onf}. The origin of nonzero neutrino masses is currently unknown, but it is clear that new degrees of freedom must exist. The preferred neutrino mass mechanism makes use of the fact that neutrinos are singlets of the unbroken SM gauge groups and hence can have Majorana masses. In the simplest realizations, the so called Seesaw Mechanisms  \cite{typeI, typeII, typeIII}, a heavy state is introduced that mediates the Weinberg operator, $(LH)(LH)$. On the other hand, Dirac neutrino masses also require the introduction of a new, SM gauge-singlet fermionic degree of freedom. While technically natural,  this hypothesis is often considered to be contrived since it requires extremely small Yukawa couplings. 

New degrees of freedom are also required in order to explain the Dark Matter (DM) puzzle. Regardless of its nature, the existence of DM also appears to imply a new fundamental physics scale (see \cite{deGouvea:2012hc} for an attempt at a counter example).

In this paper, we propose that all \emph{light} fermionic degrees of freedom are chiral in the sense described above -- what we mean by ``light'' will be made clear below. In particular, new light fermionic degrees of freedom required to explain neutrino masses and DM must also be chiral. To achieve this, we postulate the existence of, at least, a new nonanomalous $U(1)$ gauge symmetry that we dub $U(1)_\nu$. We impose, similar to the SM, that (i) all new fermions are charged under $U(1)_{\nu}$ -- that plays the role of SM hypercharge -- and (ii) no light vector-like fermions exist. Anomaly cancellations then require the existence of several new fields. With these ingredients we construct a model for a Dark Sector (DS) that, in analogy to the SM, shows accidental global symmetries as a consequence of the $U(1)_\nu$ charge assignments. These in turn imply the existence of at least one stable massive particle that serves as a DM candidate.

Similar to the SM active neutrino, massless fermions appear in the DS. We show that the smallness of neutrino masses can be understood if the initially massless DS states are charged under the SM lepton number symmetry. Lepton number is communicated between the SM and the DS via a nonchiral, heavy mediator sector charged under both the SM and the DS gauge symmetries. After integrating out the mediator sector, this model produces naturally small Dirac neutrino masses through a mechanism first described in \cite{Roncadelli}. This contrasts with previous models of a chiral DS, e.g. \cite{HeeckZhang}, where the Seesaw is still invoked.

Kinetic mixing between the SM and the $U(1)_\nu$ gauge bosons \cite{kineticmix}  is unavoidable in this scenario. It leads to several consequences, most importantly, the possibility of detecting the DM particles in the laboratory \cite{DMkinmix}. Barring the possibility of fine tuning, the predicted DM cross section would be within reach for the next generation of DM direct detection experiments.

In order to develop the scenario described above, an algorithm is necessary to write chiral models that are at least anomaly-free under a $U(1)$ gauge symmetry \cite{DobrescuSpivak} -- see also \cite{Yanagida, MuChun}. A general technique to do so is described in Sec.~\ref{so10}, towards the end of this manuscript. Indeed, following the steps detailed in Sec.~\ref{so10}, one can construct any number of models that fit our requirements. The method also allows one to address several technical questions including what are the ``minimal'' anomaly-free $U(1)$ gauge theories with chiral fermions.  This question is answered precisely, for two different criteria for minimality: smallest highest charge and smallest number of fermion fields.

Before embarking in such general considerations, we describe in Section~\ref{model} a simple yet phenomenologically appealing realization of our scenario and, 
in Section~\ref{new_u1}, discuss in some detail some of the relevant features of this model, mostly those related to the introduction of new gauge interactions.
We discuss how small neutrino masses and dark matter can be accommodated in this model in Sections~\ref{nu_mass} and \ref{DM}.  
Finally, in Section~\ref{end}, we briefly discuss other models, potential research directions, and provide some concluding thoughts.


\section{Chiral $U(1)_{\nu}$ Model}
\label{model}


In this section we construct a Lagrangian consisting of the SM, a $U(1)_\nu$-charged, nonanomalous ``dark sector (DS)'' and a ``mediator sector''. 
\be
\msc{L} = \msc{L}_{\rm SM} + \msc{L}_{\rm DS} + \msc{L}_{\trm{Mix}} + \msc{L}_{\rm Med},
\label{eq:model}
\ee
where $ \msc{L}_{\rm SM}$, $\msc{L}_{\rm DS}$, and  $\msc{L}_{\rm Med}$ represent the SM, DS, and Mediator Lagrangians, respectively. The term $\msc{L}_{\trm{Mix}}$ contains renormalizable operators that ``mix'' the SM and DS degrees of freedom, including the kinetic mixing of $U(1)_{\nu}$ with the hypercharge $U(1)_Y$ and the scalar potential coupling between the Higgs field and the equivalent ``dark Higgs'' scalar field. We discuss these term in detail in Sec.~\ref{new_u1}.

In addition to the SM gauge group, $\msc{L}$ is also invariant under a gauged $U(1)_{\nu}$ symmetry, which ``lives'' in the DS. We assume the fermionic particle content, assuming all fermions to be chiral, is as follows, keeping in mind that all fermions are left-handed Weyl fields: 
\begin{itemize}
\item three fields with charge $+1$ -- $\mbf{1}_+^{0,1,2}$;
\item two fields with charge $-2$ -- $\mbf{2}_-^{1,2}$;
\item two fields with charge $-3$ -- $\mbf{3}_-^{1,2}$; 
\item three fields with charge $+4$ -- $\mbf{4}_+^{0,1,2}$; 
\item  one field with charge $-5$ -- $\mbf{5}_-^0$. 
\end{itemize}
We adopt a normalization for the new gauge coupling $g_{\nu}$ in which all  $U(1)_{\nu}$ charges are integers. The SM fields are not charged under $U(1)_{\nu}$, while the new fermions defined above are not charged under the SM gauge symmetry.\footnote{From the point of view of the SM, all new fermions are gauge-singlet ``neutrinos,'' hence the name $U(1)_{\nu}$.} $U(1)_{\nu}$ is assumed to be spontaneously broken in order to render most of the new fermions and the new gauge boson massive. We achieve this by, similar to the SM, adding one scalar field $\phi$ with charge $+1$ and a scalar potential such that $\phi$ has a nonzero vacuum expectation value $v_{\phi}$. 

Having defined the gauge symmetry and the particle content, the renormalizable Dark Sector Lagrangian is well-defined:
\be
\msc{L}_{\trm{DS}} = \msc{L}_{\trm{DS-kin}} +  \msc{L}_{\trm{DS-Yuk}} + V(\phi)
\ee
where $\msc{L}_{\trm{DS-kin}}$ and $\msc{L}_{\trm{DS-Yuk}}$ represent the kinetic-energy and Yukawa terms respectively, while $V(\phi)$ is the scalar potential for the field $\phi$. The kinetic-energy terms  are 
\be
\msc{L}_{\rm DS-kin} =  -\frac{1}{4}\tilde{B}_{\mu\nu}\tilde{B}^{\mu\nu} + i\bar{\mbf{1}}_+^i \bar{\sigma}_{\mu}D^{\mu}_{+1} \mbf{1}_+^i + i\bar{\mbf{2}}_-^k \bar{\sigma}_{\mu}D^{\mu}_{-2} \mbf{2}_-^k + i\bar{\mbf{3}}_-^k \bar{\sigma}_{\mu}D^{\mu}_{-3} \mbf{3}^k_- + i\bar{\mbf{4}}_{+}^i \bar{\sigma}_{\mu}D^{\mu}_{+4} \mbf{4}_+^i + i\bar{\mbf{5}}^0_- \bar{\sigma}_{\mu}D^{\mu}_{-5} \mbf{5}^0_- + |D^{\mu}_{+1}\phi|^2  \label{DS-kin}
\ee
where
\be
D^{\mu}_{q} = \partial^{\mu} - i g_\nu q \tilde{B}^{\mu},
\ee
$\tilde{B}_{\mu\nu}$ is the $U(1)_{\nu}$ field strength and $\tilde{B}^\mu$ the $U(1)_{\nu}$ gauge field.  $i = 0,\,1,\,2$, $k = 1,\,2$. Note that, here, $\bar{\mbf{1}}_+$ (which has charge $-1$) represents the complex-conjugated, right-handed Weyl field associate to the $\mbf{1}_+$ left-handed Weyl field, etc. 

The Yukawa interactions in the DS are given by:
\be
-\msc{L}_{\rm DS-Yuk} =  f_{ik}\mbf{1}_+^i\mbf{2}_-^k\phi  +  h_{i0}\,\mbf{4}_+^{i}\mbf{5}_-^0\phi + h_{ik}\,\mbf{4}_+^{i}\mbf{3}^k_- \phi^*  + \trm{h.c.}, \label{DS-Yuk}
\ee
where $f$, $h$ are $2\times 3$ and $3\times 3$ Yukawa matrices respectively. In what follows the indices $i, \,k$ will be omitted whenever there can be no confusion.

After $U(1)_{\nu}$ breaking, the $\mbf{3}_-$,  $\mbf{5}_-$ and $\mbf{4}_+$ fields ``pair up'' into three Dirac fermions, labelled from here on $\chi^{i}$, their masses-squared associated to the eigenvalues of the $3\times 3$ mass-squared matrix $M_\chi M_\chi^{\dagger}$, where 
\begin{equation}
(M_\chi)_{ij} = h_{ij} v_\phi\,,\quad i,\,j=0,\,1,\,2.
\end{equation} 
In general, $M_\chi$ is expected to have three nonvanishing eigenvalues and hence describe three massive Dirac fermions.

Similarly, the two $\mbf{2}_-$ fields pair up with two linear combinations of the three $\mbf{1}_+$ fields into two Dirac fermions that we label $N^k$. Note that there is no $\mbf{2}_-^0$ field. Hence, there is a massless chiral linear combination of $\mbf{1}_+$ fields, that we call $\nu^c$.  All the masses-squared, including the vanishing one, can be obtained from the eigenvalues of the $3\times 3$ mass-squared matrix $M_{\nu N}M_{\nu N}^{\dagger}$, where 
\begin{equation}
(M_{\nu N})_{ij} = f_{ij} v_\phi\,,\quad i,\,j=0,\,1,\,2\,,
\label{MnuN}
\end{equation} 
and where we define $f_{i0} \equiv 0$ for all $i=0,1,2$.  In summary, in the absence of fields that connect the DS to the SM, after symmetry breaking, the model contains $3+2=5$ massive Dirac fermions $N^{1,2}$, $\chi^{1,2,3}$, and one massless left-handed Weyl fermion, $\nu^c$.

The unitary rotation of the fields $\mbf{1}_+$ and  $\mbf{2}_-$ that renders $M_{\nu N}$ diagonal has no effect in Eq.~\eqref{DS-kin} since these rotations operate independently on fields of the same charge. Performing a similar change of basis on the $\mbf{3}_-$, $\mbf{4}_+$ and $\mbf{5}_-$ fields in order to render $M_\chi$ diagonal is different since, in this case, the $\mbf{5}_-^0$ and $\mbf{3}_-^{1,2}$ fields must be grouped together as components of a 3-vector. If $M_\chi$ is diagonalized by the transformation
\be
\mcl{U} M_\chi \mcl{V}^\dagger = \trm{diag}\{ M_{\chi_1}, M_{\chi_2}, M_{\chi_3}\} \,,
\ee
where $\mcl{U,V}$ are $3\times 3$ unitary matrices, ``mixing'' appears among the $\chi$ fields. The $U(1)_{\nu}$ couplings to the $\chi_i$ mass eigenstates are off-diagonal which indicates the presence of ``flavor-changing neutral currents'' within each DS generation. Explicitly, in the mass basis, the $\tilde{B}$ couplings are given by\footnote{The Dirac fermions are (ignoring $i,k$ indexes), $N^T=\left(\mbf{2}_-, \bar{\mbf{1}}_+\right)$, $\chi^T=\left(\mbf{3}_-,\bar{\mbf{4}}_{+}\right),\left(\mbf{5}_-,\bar{\mbf{4}}_{+}\right)$, where the comma indicates that $\chi$ are linear combinations of those two Dirac fermions.}
\be
 g_{\nu}\left( \bar{\nu^c} \bar{\sigma}_{\mu} \nu^c - \bar{N}_R \gamma_\mu N_R - 2\bar{N}_L \gamma_\mu N_L - 4\bar{\chi}_R \gamma_\mu \chi_R +  \bar{\chi}_L \mcl{Q}_{35}\gamma_\mu \chi_L \right) \tilde{B}^{\mu}, \label{DS-kin2}
\ee
where the $L$, $R$ subindices mean that we take the left- or right-handed components of the corresponding Dirac field. Indices running over fields of the same charge are implied and the matrix $\mcl{Q}_{35}$ in the last term is the rotated charge-matrix
\be
\mcl{Q}_{35} = \mcl{V} \left( \begin{array}{ccc}
  -3 && \\ & -3 & \\ && -5
\end{array} \right) \mcl{V}^\dagger \,. \label{Q35}
\ee
Since this matrix is nondiagonal, generically, two of the $\chi$ fermions in the 345-sector, say $\chi_2$ and $\chi_3$, are unstable; decaying for example into a lighter $\chi$ field and a $\nu^c\bar{\nu^c}$ pair via $\tilde{B}$ boson exchange. This $\tilde{B}$ boson can be on- or off-shell depending on the masses of the $\chi_i$. Also notice that the stable, lightest $\chi$-particle, $\chi_1$, couples to $\tilde{B}$ with a right-handed coupling $-4g_{\nu}$ and a left-handed coupling $(-3+(-5+3)|\mcl{V}_{\chi 5}|^2)g_{\nu}$, where $|\mcl{V}_{\chi 5}|^2$ is the probability that a $\chi_1$ state will interact as a charge $-5$ object. The scattering cross-section of an unpolarized $\chi_1$ beam via $\tilde{B}$ exchange will be proportional to an effective charge-squared
\be
Q_{\chi}^2=\frac{1}{2}\left(4^2 + \left[(3+2|\mcl{V}_{\chi 5}|^2\right]^2\right)\in [12.5,17], \label{Qchi1}
\ee
which will come in handy later. 

After all the smoke has cleared, $\msc{L}_{\rm DS}$ contains four accidental global $U(1)$ symmetries, $U(1)_{\chi} \times U(1)_{N^1}\times U(1)_{N^2}\times U(1)_{\nu^c}$  and hence four stable particles. More structure in the DS will, in general, reduce the accidental global symmetry of the Lagrangian. For example, if there is a second $U(1)_{\nu}$ charged scalar field $\phi'$, also with charge $+1$, the $U(1)^4$ global symmetry would be, in general, reduced to $U(1)_{\chi}\times U(1)_{12}$, where $12$ refers to the fields with charges $+1,-2$. In this case, only the lightest $\chi$ field would be absolutely stable, along with the massless $\nu^c$ state. Ultimately, we will associate the lightest $\chi$ field with dark matter, while $\nu^c$ will play the role of the left-handed antineutrino SM gauge singlet field.  

Before proceeding, it is interesting to establish a parallel between the DS and the SM. In the SM, assuming only one generation and ``turning off'' the strong and the charged-current weak interactions, there are 7 massive Dirac fermions ($u\times 3$, $d\times 3$, $e$) and one massless left-handed Weyl fermion (the neutrino component of $L$, $\nu_L$). There is also a very large accidental classical global symmetry, $U(1)^8$ (``green up-quark number,'' ``electron number,'' ``neutrino number,'' ``red down-quark number, etc). The charged-current weak interactions explicitly break this down to $U(1)^3_q\times U(1)_\ell$ for quarks (one for each color) and leptons, respectively.

In summary, the chiral $U(1)_{\nu}$ model described here accommodates new stable massive fermions whose masses are proportional to a new mass scale, which we will associate with the dark matter. As we will discuss in Sec.~\ref{DM}, the new gauge interaction, combined with the interactions between the DS and the SM discussed in the next section, is sufficient to predict the dark matter relic density. In order to render the neutrinos massive, however, one is required to add new mediator fields charged under both the SM and $U(1)_{\nu}$, and a new mass scale, described in $\msc{L}_{\rm Med}$. We deal with this issue in Sec.~\ref{nu_mass}


\section{The mediator sector, $U(1)_{\nu}$ interactions and kinetic mixing}
\label{new_u1}


We postulate the existence of heavy nonchiral degrees of freedom charged under the gauge symmetries of both the SM and the DS. For concreteness, let us choose a very simple possibility: one vector-like Dirac fermion $X$ with mass $\Lambda$. $X$ is a weak doublet with hypercharge $+1/2$ and $U(1)_{\nu}$-charge $-1$ whose renormalizable Lagrangian reads (we define $X^T\equiv(x,\bar{x}^c)$, where $x,x^c$ are left-handed Weyl fermions):
\begin{equation}
  \msc{L}_{\rm Med} = i\bar{X}\slashed{D}X + \Lambda \bar{X}X - \left( \kappa_L Lx\phi + \kappa_i
  \mbf{1}_+^i x \tilde{H}  + \trm{h.c.} \right) \,, \label{med-lag}
\end{equation}
where
\be
\slashed{D} = \slashed{\partial} - i\frac{g}{2}\slashed{W}^aT^a - i\frac{g'}{2}\slashed{B} + ig_\nu \slashed{\tilde{B}} \,,
\ee
$T^a$ are the Pauli matrices, and $L^T = (\nu_L, \ell_L)$ is the SM lepton doublet. We will discuss most of the consequences of this part of the Lagrangian in the next section. Before doing that, we need to address the new gauge interactions. 

The fact that $\msc{L}$ is invariant under $U(1)_Y\times U(1)_{\nu}$ implies that one must also take kinetic mixing between the two $U(1)$ field strengths into account:
\be
\msc{L}_{\trm{Kin-Mix}} = -\frac{\sin \eta}{2} B^{\mu\nu} \tilde{B}_{\mu\nu},
\ee
where $B^{\mu\nu}$ represents the hypercharge field strength. In the next sections  we will argue that phenomenological considerations impose a bound $g_\nu \sin\eta \lesssim 10^{-3}$. We now show that this value is natural in the sense that quantum corrections do not destabilize it.

If the kinetic mixing is set to zero at some scale, quantum corrections give rise to a nonvanishing $\eta$. These effects appear as a consequence of the mediator sector, which couples to both the SM and the DS $U(1)$s, and generates a nonzero $\sin\eta$ at the one-loop level. This one-loop correction to $\sin\eta$ is given by
\be
\Delta\sin\eta = C_\eta\frac{g'g_\nu}{16\pi^2}\log\left(\frac{\Lambda}{\mu}\right), \label{sineta}
\ee
where $g'$ is the hypercharge gauge coupling, $\mu$ is the renormalization scale and $C_\eta$ is an $\mcl{O}(1)$ coefficient that depends on the renormalization conditions. Taking for the combination $C_\eta g'g_\nu \sim 0.1$, Eq.~\eqref{sineta} implies that $\sin\eta \sim 10^{-3}$ is natural.

After electroweak and $U(1)_\nu$ symmetry breaking, the physical gauge bosons $A_\mu$, $Z_\mu$ and $\tilde{Z}_\mu$ are related to $B_\mu$, $W^3_\mu$, $\tilde{B}_\mu$ by the following nonunitary linear transformation \cite{Babu:1997st}: 
\be
\left( \begin{array}{c} B_\mu \\ W^3_\mu \\ \tilde{B}_\mu \end{array} \right) = \left( \begin{array}{ccc}
  c_W & -\cos\xi s_W - \tan\eta \sin\xi & \sin\xi s_W - \tan\eta\cos\xi \\
  s_W & \cos\xi c_W & -\sin\xi c_W \\
  0 & \sin\xi \sec\eta & \cos\xi \sec\eta
\end{array} \right) \left( \begin{array}{c} A_\mu \\ Z_\mu \\ \tilde{Z}_\mu \end{array} \right) \,,
\ee
where $c_W$ and $s_W$ are, respectively, the sine and cosine of the weak mixing angle, while the angle $\xi$ can be succintly written in the limit $\eta\rightarrow 0$, $M_{Z} \ll M_{\tilde{Z}}$ as \cite{Babu:1997st}
\be
\sin\xi \simeq -\frac{M_Z^2}{M_{\tilde{Z}}^2} s_W \sin\eta \,. \label{sinxi}
\ee
Unless otherwise noted, we will mostly be interested in the limit in which Eq.~\eqref{sinxi} holds. Notice that in this limit  $\xi \ll \eta$.

In general, we find that DS particles acquire an ${\cal O}(\xi)$ small coupling to the physical $Z$ boson while SM particles acquire an ${\cal O}(\eta)$ coupling to the physical $\tilde{Z}$ gauge boson. These couplings are important because they represent the main SM--DS interaction and can be potentially probed experimentally.  On the other hand, no coupling of the DS to the photon appears as it is expected since electromagnetism remains unbroken. Explicitly, the couplings of the DS particles to the $Z$ are
\be
  g_\nu \sin\xi \left( \bar{\nu^c} \bar{\sigma}_{\mu}\nu^c  - \bar{N}_R \gamma^\mu N_R - 2\bar{N}_L \gamma^\mu N_L - 4 \bar{\chi}_R \gamma^\mu \chi_R + \bar{\chi}_L \mcl{Q}_{35} \gamma^\mu \chi_L \right)   Z_{\mu}\label{DM-Z-int} \,.
\ee

The SM particles also interact with the physical $\tilde{Z}$. In the same approximation as before -- $\eta \ll 1$, $M_{Z} \ll M_{\tilde{Z}}$ -- the interaction can be written as
\be
e\,\sin\xi \tilde{Z}_\mu   \sum_f \left[  \tan\theta_W  \left( 1  + \frac{M^2_{\tilde{Z}}}{s_W^2M_Z^2} \right)Y^{(f)} + \frac{I_3^{(f)}}{2}\cot\theta_W  \right] \bar{f}\gamma^\mu f \label{SM-Z'-int} \simeq - \frac{e\sin\eta}{c_W}\tilde{Z}_\mu \sum_fY^{(f)}\bar{f}\gamma^\mu f \,,
\ee
where the sum runs over all chiral SM fermions and
$I_3^{(f)}$ and $Y^{(f)}$ are the weak isospin and hypercharge of the fermion field $f=u_L,u_R,d_L,d_R,\ldots$. For $M_Z^2 \ll M_{\tilde{Z}}^2$ the term proportional to $Y^{(f)}$ dominates.

There is also mixing in the scalar sector. Again, assuming that the coefficient $\lambda_{\phi H}$ of the $\lambda_{\phi H} |\phi|^2|H|^2$ term vanishes at tree level, a coupling $\lambda_{\phi H}$ would be generated at the one-loop level of order
\be
\lambda_{\phi H} \sim \frac{\kappa^2_L(\sum_i\kappa^2_i)}{16\pi^2}\log\left(\frac{\Lambda}{\mu}\right), \label{lambdaphiH}
\ee 
where we ignore a $Z-\tilde{B}$ one-loop diagram proportional to $\sin^2\xi$ (which can be considered a three-loop effect). We take the value in Eq.~\eqref{lambdaphiH} to be the ``natural'' value for $\lambda_{\phi H}$. For small $\kappa_L$ and $\kappa_i$, $\lambda_{\phi H}$ is naively more suppressed than $\sin\eta$. Moreover, as we will briefly argue in the concluding section, if one were interested in avoiding large finite corrections from the mediator scale $\Lambda$ to the scalar masses-squared, one would be forced to impose $\kappa_L,\kappa_i\lesssim 10^{-4}$. Henceforth, we will assume for simpliciy that kinetic mixing effects are more significant than those related to scalar mixing.


\section{Neutrino Masses}
\label{nu_mass}


The mediator Lagrangian in Eq.~\eqref{med-lag}  explicitly breaks the $U(1)_{N^1}\times U(1)_{N^2}\times U(1)_{\nu^c}\times U(1)_{\ell}$ down to $U(1)_L$ -- what is normally referred to as lepton-number. That is, the introduction of new, heavy degrees of freedom necessarily renders all neutrinos and DS particles massive, but does it in a way that lepton number remains a good quantum number \cite{Roncadelli}. Examples of recent models that make use of this idea are \cite{Diracmodels}.

Upon integrating out the $X$ field\footnote{We will mostly be interested in the physics at energy scales well below the mass of these mediator fields and could also introduce their effects by adding higher-dimensional operators to $\msc{L}_{\rm SM} + \msc{L}_{\rm DS}$. We find that the introduction of a concrete model renders the discussion more transparent.}, one generates the following dimension-five effective operator, assuming only one generation of SM leptons:
\begin{equation}
\frac{\kappa_L\kappa_i}{\Lambda} \left(\mbf{1}_+^iL\right)\left(H\phi\right) + \trm{h.c.}.
\label{eq:dim-5}
\end{equation}

It is important to note that the mediator sector we have chosen is an example. Any UV-completion that implements the symmetry breaking pattern described above would also lead to the operator in Eq.~\eqref{eq:dim-5}. After symmetry breaking, this operator manifests itself as a Dirac mass between the $\mbf{1}_+$ fields and $\nu_L$, 
\be
m^D_i = \frac{\kappa_L\kappa_i v v_\phi}{\Lambda}.
\label{small_m}
\ee
where $v$ is the vev of the Higgs field, $v \sim 10^2$ GeV. 
In more detail, after $U(1)_\nu$ symmetry breaking, the three pairs of Weyl fields $(\nu_L,\, \mbf{2}_-^{1,2})$ and $\mbf{1}_+^{0,1,2}$ combine into three massive Dirac fermions, with mass matrix
 \begin{equation}
(M_{\nu})_{ij} = f'_{ij} v_\phi,
\end{equation}  
where $f'_{ij}=f_{ij}$ for $i=0,1,2$ and $j=1,2$, while $f'_{i0}\equiv m^D_i$ for $i=0,1,2$. In the basis where $M_{\nu N}$, defined in Eq.~(\ref{MnuN}), is diagonal,
\begin{equation}
M_{\nu}=\left(\begin{array}{ccc} m^D_0 & 0  & 0 \\ m^D_1 & M_1 & 0 \\ m^D_2 & 0 & M_2 \end{array}\right),
\end{equation}
where $M_1$, $M_2$ are the eigenvalues of $M_{\nu N}$, naively of order $v_\phi$. The masses $m^D$, on the other hand, are parametrically smaller than $M_{1,2}$ by a factor $v/\Lambda$. In the limit $M_{1,2}\gg m^D_{0,1,2}$, the eigenvalues of $M_{\nu}$ are simply $m^D_0$, $M_1$, $M_2$. With respect to the SM weak interactions, the three massive Dirac fermions are a mostly active neutrino $\nu$ and two mostly sterile neutrinos $N_{1,2}$.

The fact that there are three generations of weakly interacting neutrinos in the SM, two of which are known to be massive, implies that the number of DS ``families'' $N_f$ is at least two. For $N_f=2$, there are three mostly active neutrinos, one of which is massless, and four mostly sterile states. All massive states are Dirac fermions. If all mostly active neutrinos are massive, $N_f\ge 3$ and there are at least six mostly sterile massive states. For $N_f>3$, there is a mismatch between the number of $\nu_L$ and $\nu^c$ states, and $N_f-3$ antineutrinos remain massless.

Experimentally, neutrino masses are known to be of order $10^{-1}$~eV or less which implies
\be
\frac{\kappa_L\kappa_0 v_\phi }{\Lambda} \lesssim 10^{-11}\,.
\ee
The smallness of the neutrino masses can be attributed to $\Lambda\gg v_\phi$ or $\kappa_L\kappa_0\ll 1$. Unless otherwise noted, we will assume that $v_\phi \sim v \sim 10^2$~GeV so if $\kappa_0\kappa_L\sim 1$, $\Lambda\sim 10^{13}$~GeV is required in order to ``explain'' the small neutrino masses. On the other hand, if $\Lambda\sim 1$~TeV, small neutrino masses can be ``explained'' if $\kappa_0\kappa_L\sim 10^{-10}$. This is, qualitatively speaking, not different from the standard Type-I seesaw mechanism \cite{typeI}. It is amusing to note that, in the limit $\kappa_L=\kappa_0$ and $v=v_{\phi}$, the expression for small Dirac neutrino masses we obtain here, Eq.~(\ref{small_m}), is identical to the one for small Majorana neutrino masses from the Type-I seesaw if we identify $\Lambda$ with the right-handed neutrino masses and $\kappa$ with the neutrino Yukawa coupling. 

It is easy to diagonalize $M^\nu$\footnote{We restrict the discussion to one generation of DS and SM fields. The extension to three SM families and $N_f$ DS families is straightforward.} in the limit $M_1,M_2\gg m^D_0,m^D_1, m^D_2$. $M^{\nu}=\mcl{U}_RM^{\nu}_{\rm diag}\mcl{U}_L^{\dagger}$, where 
\begin{eqnarray}
& \mcl{U}_L = \left( \begin{array}{ccc}
  1 & m^D_1/M_1 &  m^D_2/M_2 \\
  -m^D_1/M_1 & 1 & 0 \\
  -m^D_2/M_2 & 0  &  1 
\end{array}
\right) + {\cal O}\left(\frac{m^D_i}{M_k}\right)^2, \, \\
& \mcl{U}_R = \left( \begin{array}{ccc}
  1 & 0 & 0 \\
  0 & 1 & 0 \\
  0 & 0  &  1 
\end{array}
\right) + {\cal O}\left(\frac{m^D_i}{M_k}\right)^2, \, \\
 & M^{\nu}_{\rm diag} = \left( \begin{array}{ccc}
  m^D_0 & 0 & 0 \\
  0 & M_1 & 0 \\
  0 & 0  &  M_2 
\end{array}
\right) \, .
\end{eqnarray}
In the case of $M^{\nu}_{\rm diag}$, we also only keep the leading order $m^D_i/M_k$ terms.  In the mass basis, the gauge interactions will couple the heavy, sterile states $N$ to the active $\nu$ state. In more detail, the following interactions appear after diagonalization of $M_\nu$:
\be
-\frac{m^D_1}{M_1}\bar{\nu}_L\left( \frac{g}{2c_W}\slashed{Z} - 2g_\nu \tilde{\slashed{Z}} \right) N_{1L} - \frac{m^D_2}{M_2}\bar{\nu}_L\left( \frac{g}{2c_W}\slashed{Z} - 2g_\nu \tilde{\slashed{Z}}\right) N_{2L} + \trm{h.c.} \,,
\ee
describing the potential decays $N_k \rightarrow Z \nu$, $\tilde{B}\nu$. Moreover,  from the standard weak couplings to the $W$-boson, we find
\be
\frac{g}{\sqrt{2}} \, \left( -\frac{m^D_1}{M_1}\bar{\ell}_L \slashed{W}^- N_{1 L}  - \frac{m_2}{M^D_2}\bar{\ell}_L \slashed{W}^- N_{2 L} \right) + \trm{h.c.}\,,
\ee
which allow for the interesting decays of the DS $N$ particles into charged leptons, $N_k\to \ell W^{(*)}$.
The strength of the coupling of $N$s to the SM gauge bosons, proportional to the ratio between the neutrino masses and the masses of the sterile neutrinos, is a generic feature of these models.

In the case in which the $N$ particles are heavier than the weak bosons, the decay rate of, say, $N_1\rightarrow \ell W$ is given by
\be
\Gamma_{N_1\rightarrow \ell W} = \frac{G_F (m^D_1)^2M_1}{8\pi\sqrt{2}} + {\cal O}\left( \frac{M_W^2}{M_1^2} \right)
\ee
from which we obtain $\Gamma_{N_1\rightarrow \ell W}\sim 0.4\trm{ s}^{-1}$ for $m^D_1 = 0.1$ eV, $M_1 = 1$~TeV. On the other hand, if the SM gauge bosons are heavier than the $N_i$ fields, these decay to SM fields via off-shell $W$-bosons and $Z$-bosons. In this case, the decay width of $N\to SM$ scales like $M_i^3$ -- a factor of $M_i^5$ from kinematics times the ``mixing parameter'' squared, proportional to $(m^D_i/M_i)^2$. For very light sterile neutrino masses -- masses below 1~MeV --  at the tree-level, only the $N\to \nu\bar{\nu}\nu$ decays are kinematically available, and the one-loop suppressed decay $N\to \nu\gamma$ also becomes relevant. The couplings associated to these decays are again proportional to $m^D/M$. In summary, assuming all $m^D_i$ are of order the active neutrino masses, the sterile neutrino lifetimes range between tenths of milliseconds for $M_{N}\sim 1$~10 TeV to order $10^5$~years for $M_{N}\sim 100$~MeV and much longer than the age of the universe for lighter sterile neutrinos.

It is intriguing that for sterile neutrino masses of order 10~keV, the active-sterile mixing angle squared $(m^D/M)^2$ is of order $10^{-10}$, in agreement with the recent ``sterile-neutrino-as-dark-matter'' interpretations of the 3.5 keV line \cite{Bulbul:2014sua}. Here, however, one needs to revisit the issue with some care since, in the early universe, the sterile neutrinos are kept in thermal equilibrium with the photons via flavor-diagonal $Z$ and $\tilde{Z}$ interactions, discussed in some detail in the previous section. These interactions determine their relic abundance, as opposed to the standard Dodelson-Widrow mechanism \cite{Dodelson:1993je}, where active-sterile mixing determines the relic abundance of the mostly sterile states (for other possibilities see, for example, \cite{Abazajian:2014gza}). We discuss early-universe related issues in more detail in Sec.~\ref{DM}.  

Sterile neutrinos can be produced in the laboratory mostly via their coupling to the $Z$ and the $\tilde{Z}$ ($Z^{(*)},\tilde{Z}^{(*)}\to N_i\bar{N}_i$), or the scalar fields associated to spontaneous symmetry breaking. The associated phenomenology and potential current constraints will depend on the mass of the sterile neutrinos but, for most masses, as discussed above, the $N_i$ particles are effectively stable when compared to the time-scales of laboratory experiments and will manifest themselves as missing energy.


\section{Early Universe Cosmology and Dark Matter}
\label{DM}


Here we provide a brief discussion of the thermal history of the universe described by the Lagrangian in Eq.~(\ref{eq:model}). As discussed in Sec.~\ref{model}, in the scenario under investigation there are a number of accidental global symmetries that ensure the presence of stable particles. If there were no mediator states $X$, for every generation of new fermions, the corresponding $N_1$, $N_2$ would be stable, along with the massless ``left-handed antineutrinos'' $\nu^c$. Furthermore, $\chi_1$, the lightest of the $\chi_i$ fields, is also stable. The mediator interactions discussed above, other than rendering the $N_{1,2}$ unstable on cosmological time-scales do not play a significant role at temperatures much smaller than $\Lambda$. 
In what follows, we estimate the current constraints on the DS from cosmological measurements, and discuss whether the $\chi$ particles can explain the dark matter puzzle. For concreteness, we will use the following values for the DS parameters in our estimates:
\be
M_1 \sim M_2 \sim \trm{1 TeV}\,, \quad M_\chi \sim \trm{5 TeV} \,, \quad M_{\tilde{Z}} \sim \trm{500 GeV} \,, \quad g_\nu \sim 0.1 \,. \label{ex-values}
\ee

We denote the temperatures of the SM and DS plasmas by $T$ and $\tilde{T}$ respectively. The distinction is only relevant below a certain temperature $T_{\rm dec}$ at which the interactions that couple the two sectors fall out of thermal equilibrium. Above $T_{\rm dec}$ there is a single plasma composed of SM and DS states. Also, let $\mbb{g}^*$ represent the effective number of relativistic degrees of freedom at temperatures greater than $T_{\rm dec}$ while $g^*$ and $\tilde{g}^*$ represent the degrees of freedom below $T_{\rm dec}$  for the SM and DS plasmas respectively. To simplify things we assume that no reheating takes place at the time of decoupling so that at $T_{\rm dec}$ we have the matching condition:
\be
\mathbb{g}^*_{\rm dec} = g^*_{\rm dec} + \tilde{g}^*_{\rm dec} \,.
\ee

The thermal history of this model proceeds as follows \cite{ChackoCui}. As the universe cools down, DM becomes nonrelativistic and freezes out  eventually, leaving a thermal relic of $\chi_1$ particles. Annihilations of the heavy, stable $\chi_1$ particles into lighter $\nu^c$ or $N_i$ dominate over annihilations into SM states and are the main processes that determine the relic density. Shortly after freeze-out, all the states in the DS are nonrelativistic except for the $\nu^c$.\footnote{For masses of the sterile neutrinos $M_1$, $M_2 \sim 1$ TeV,  their lifetimes are a fraction of a second and hence ``safe'' with respect to  constraints from big-bang nucleosynthesis, since the injected particles have sufficient time to thermalize. We assume therefore that the sterile neutrinos are mostly harmless as far as cosmological observables are concerned and ignore them henceforth, unless otherwise noted. }  These remain coupled to the SM plasma up to $T_{\rm dec}$. 
From this point on, $T$ is no longer necessarily equal to $\tilde{T}$ and the two gases must be treated independently. After the time of decoupling, adiabaticity of the evolution of the two sectors then imposes the constraint
\be
\frac{\tilde{g}^*\tilde{T}^3}{g^*T^3} = \frac{\tilde{g}^* _{\rm dec}}{g^*_{\rm dec}} \label{Tratios} \,.
\ee

We start by estimating $T_{\rm dec}$ and how it is constrained by observations. Given the choice of parameters Eq.~(\ref{ex-values}), this occurs at temperatures well below the electroweak and $U(1)_{\nu}$ phase transitions. 
At temperatures $T = \tilde{T} \lesssim 100$ GeV, the only relativistic state in the DS plasma are the $\nu^c$. The presence of new, light particles in thermal equilibrium with the SM gas is contrained by a variety of observations. 
This is true of the antineutrino fields $\nu^c$, which are effectively massless. These degrees of freedom persist throughout the thermal history of the universe and contribute to its expansion rate, especially when the universe is radiation dominated. More concretely, they contribute to the parameter $N_{\rm eff}$, which parameterizes the number of relativistic degrees of freedom:
\be
N_{\rm eff} = N_{\rm eff}^{\rm SM} + \Delta N_{\rm eff}\,,\quad \Delta N_{\rm eff} = \frac{\rho_{\rm eR}}{\rho_{\nu}} \,.
\ee
Here $N_{\rm eff}^{\rm SM}$ is the SM contribution while $\Delta N_{\rm eff}$ comes from new physics and is defined as the ratio between the energy density of extra radiation $\rho_{\rm eR}$ and that of one SM neutrino $\rho_{\nu}$.

The SM particle content and interactions translate into $N_{\rm eff}^{\rm SM} = 3.045$. Measurements constrain  $N_{\rm eff}$ at the time of big-bang nucleosynthesis (BBN) and at the surface of last scattering. In particular, if the gas of $\nu^c$ states were at the same temperature as the active neutrino gas at the time of BBN, they would contribute $\Delta N_{\rm eff}=N_{\rm f}$. Since we need at least two DS generations in order to account for two massive neutrinos, we have $N_{\rm f} \ge 2$, a possibility that would be in tension with the current bounds if it were to translate directly into $\Delta N_{\rm eff}$. Measurements of $\Delta N_{\rm eff}$ come from BBN and the cosmic microwave background (CMB). The Planck collaboration has recently published values $N_{\rm eff} = 3.15 \pm 0.46 $ at 95\% C.L. \cite{Ade:2015xua} while in \cite{BBNbounds1} the bound $\Delta N_{\rm eff}^{\rm BBN} < 1.5 $ at the time of BBN was found.

On the other hand, if the DS gas decouples early enough from the SM its temperature $\tilde{T}$ at the time of big bang nucleosynthesis can be significantly lower than that of the active neutrinos $T$. If that is the case,
\begin{equation}
\Delta N_{\rm eff}  = \frac{4 \tilde{g}^*}{7} \frac{\tilde{T}^4}{T^4} = \frac{4 \tilde{g}^*}{7} \left( \frac{g^*\, \tilde{g}^*_{\rm dec}}{g^*_{\rm dec} \, \tilde{g}^*} \right)^{4/3}   \,,
\end{equation}
where we have used Eq.~\eqref{Tratios}. Assuming $N_f = 3$ generations in the dark sector $\tilde{g}^* = \tilde{g}^*_{\rm dec} = 6$, and taking for $g^*$ its value at the time of neutrino decoupling $g^* = 10.75$, we find that in order to satisfy $\Delta N_{\rm eff} < 1$, it is enough to have $g^*_{\rm dec} \gtrsim 27$. Hence, it suffices that the $\nu^c$'s decouple before the QCD phase transition, which corresponds to $g^*_{\rm dec} = 61.75$, in order to satisfy the bounds comfortably. In that case we find $\Delta N_{\rm eff} = 0.33$, in agreement with Plack bounds.

For later use, it proves useful to define the ratio between the DS and SM temperatures today (indicated by the subscript `0') which has remained constant after the photon reheating by electron-positron annihilation:. Using Eq.~\eqref{Tratios} we find:
\be
r_0 \equiv \frac{T_0}{\tilde{T}_0} \lesssim 0.56 \,. \label{Tratio}
\ee

Imposing that the DS should decouple from the SM before the QCD phase transition translates into bounds on the couplings of the $\nu^c$ and SM fermionic currents to the $Z$ and $\tilde{Z}$ bosons respectively. We note that $\nu^c$ thermal equilibrium with the SM is very similar to the thermal equilibrium of active neutrinos since these also interact with the rest of the SM gas via weak interactions. Since the decoupling temperature of active neutrinos is proportional to $G_F^{-2/3}$, we roughly estimate the decoupling temperature of the $\nu^c$ from the SM gas, for $M_{Z} \ll M_{\tilde{Z}}$, as follows:
\be
\frac{T^{\rm dec}_{\nu_L}}{T^{\rm dec}} \sim \left( \frac{G_\nu G_F}{G^2_F} \right)^{1/3} \sim \left( \frac{M_{Z}g_\nu}{M_{\tilde{Z}} g} \sin\eta \right)^{2/3} \,,
\ee
where $G_\nu$ has been obtained from Eqs.~(\ref{DM-Z-int},\ref{SM-Z'-int}),
\be
G_\nu \sim \frac{\sin^2\xi\, g_\nu^2}{M_Z^2} \,.
\ee
Requiring that the $\nu^c$ decouple from the SM gas before the QCD phase transition, $T^{\rm dec}_{\nu_L}/T^{\rm dec} \lesssim 0.01$, we obtain $\sin\eta \lesssim 10^{-3}$. This matches the discussion in Sec.~\ref{new_u1}, where we argued that $\sin\eta \sim 10^{-3}$ is natural (see Eq.~(\ref{sineta})).

We proceed to estimate the DM relic density given the values of the parameters listed in  Eq.~\eqref{ex-values}.
Define as usual the time variable $x = M_{\chi_1}/T$ and let $x_{\rm fo}$ be its value at the freeze-out temperature (we will show that $T_{\rm fo} = \tilde{T}_{\rm fo}$). $x_{\rm fo}$ can be obtained from the Boltzmann equation and is typically of order
\be
x_{\rm fo} \equiv \frac{M_{\chi_1}}{T_{\rm fo}} \sim 10 \,.
\ee
It is well known that the contribution of $\chi_1$ to the energy budget of the universe today $\Omega_\chi$, is inversely proportional to its thermally-averaged annihilation cross-section, $\langle\sigma_{\rm ann} v\rangle$. For $s$-channel annihilations we have:
\be
  \langle\sigma_{\rm ann} v\rangle \sim  \sigma(\chi_1 + \bar{\chi}_1 \to \nu^c + \bar{\nu^c})\sim N_{f} Q_\chi^2 \frac{g_{\nu}^4}{8\pi M^2_{\chi_1}}\, ,
\ee
where $Q^2_\chi={\cal O}(10)$ was defined in Eq.~\eqref{Qchi1} and $N_f$ is the number of dark flavors. Putting in some numbers we find
\be
\langle\sigma_{\rm ann} v\rangle\sim 3\times 10^{-2}~{\rm pb}\times \left(\frac{1~\rm TeV}{M_{\chi_1}}\right)^2\left(\frac{N_f Q^2_\chi}{20}\right) \left(\frac{g_{\nu}}{0.1}\right)^4 \,.
\label{eq:dmcs}
\ee

For a DM number density $n_{\chi_1}$, the DM fraction today is defined, as usual, as
\begin{align}
  \Omega_{\chi 0} & = \frac{\rho_{\chi 0}}{\rho_{\rm cr}} = \frac{M_{\chi_1} Y_\infty \tilde{T}^3_0}{\rho_{\rm cr}} \left( \frac{a_\infty \tilde{T}_\infty}{a_0\tilde{T}_0} \right)^3 = \frac{M_{\chi_1} Y_\infty \tilde{T}^3_0}{\rho_{\rm cr}}  \left( \frac{a_\infty \tilde{T}_\infty}{a_{\rm dec}\tilde{T}_{\rm dec}} \right)^3  \left( \frac{a_{\rm dec} \tilde{T}_{\rm dec}}{a_{0}\tilde{T}_{0}} \right)^3 \nn \\
  & = \frac{M_{\chi_1} Y_\infty \tilde{T}^3_0}{\rho_{\rm cr}}  \left( \frac{a_\infty T_\infty}{a_{\rm dec}T_{\rm dec}} \right)^3  = \frac{M_{\chi_1} Y_\infty \tilde{T}^3_0}{\rho_{\rm cr}}  \frac{\mbb{g}^*_{\rm dec}}{\mbb{g}^*_{\infty}}  \label{Omegachi0}
\end{align}
where $Y = n_{\chi_1}/\tilde{T}^3$ and the $\infty$ subscript refers to its asymptotic value. In the above equations we have used that $T_{\rm dec} = \tilde{T}_{\rm dec}$ and $T_\infty = \tilde{T}_\infty$. Also, notice that no reheating event occurs in the DS after $\tilde{T}_{\rm dec}$ and hence $a_{\rm dec} \tilde{T}_{\rm dec} = a_0\tilde{T}_0$.

A good approximation for $Y_{\infty}$ is 
\be
Y_{\infty} \sim \frac{x_{\rm fo} H(x=1)}{M_{\chi_1}^3 \langle \sigma_{\rm ann} v\rangle} \label{Yinfty} \,,
\ee
where the Hubble rate at $x = 1$ is given by
\be
H(x = 1) = \sqrt{\frac{8\pi G \rho(x = 1)}{3}} = \sqrt{\frac{4\pi^3 G \mbb{g}^*(x = 1)}{45}} T^2  =  \sqrt{\frac{4\pi^3 G \mbb{g}^*(x = 1)}{45}} M_{\chi_1}^2 \,.
\ee
Plugging this into Eq.~\eqref{Omegachi0} we obtain 
\be
\Omega_{\chi 0} \sim  \sqrt{\frac{4\pi^3 G \mbb{g}^*(x = 1)}{45}} \frac{x_{\rm fo} T_0^3 r_0^3}{ \langle \sigma_{\rm ann} v\rangle \rho_{\rm cr}}   \frac{\mbb{g}^*_{\rm dec}}{\mbb{g}^*_{\infty}}. 
\ee
Numerically,
\be
\Omega_{\chi 0} \sim \frac{10^{-2}}{\langle \sigma_{\rm ann} v \rangle} \,  \trm{pb}.
\ee
Given the value for the thermally-averaged cross-section, estimated above (Eq.~(\ref{eq:dmcs})), $\Omega_{\chi}$ can be made to agree with the cold dark matter contribution to the energy budget of the universe,  $\Omega_ch^2=0.1188 \pm 0.0010$ \cite{Ade:2015xua}.    

The $\chi_1$ particles scatter off of ordinary matter via both $Z$ and $\tilde{Z}$ exchange, both couplings (i.e., the coupling of $\chi_1$ to the $Z$-boson and that of the SM fermions to the $\tilde{Z}$-boson) suppressed by the small kinetic mixing parameter. The cross-section for $\chi_1$--nucleus scattering is, in the limit that the dark matter is much heavier than the scattered nucleus and $M^2_{\tilde{Z}}\gg M^2_Z$~\cite{Goodman:1984dc,Cirelli:2005uq}\footnote{Even if one is interested in the limit $M^2_{\tilde{Z}}\gg M^2_Z$, $\tilde{Z}$-boson exchange is still comparable to $Z$-boson exchange in the limit where the momentum transfers are much less than $M_Z^2$, as discussed in Sec.~\ref{new_u1}.} 
\be
\sigma(\chi_1+N\to\chi_1+N)=\alpha\frac{m_N^2}{M_Z^4}\frac{g_{\nu}^2\sin^2\xi Q_V^2}{\sin^2\theta_W\cos^2\theta_W} ( 1 - \sin^2\theta_W)^2Z^2,
\ee
where $\alpha$ is the fine-structure constant, $Q^2_V=(4 + [(3+2|\mcl{V}_{\chi 5}|^2])^2$, $49 \leq Q_V^2 \leq 64$ is the square of two times the  $U(1)_{\nu}$ vector-charge of the dark matter candidate and $m_N$ is the mass of the nucleus with atomic number $Z$. It is interesting to note that in the usual WIMP scenario, the direct detection cross section is dominated by scattering off neutrons while in this case, the proton contribution is the most relevant one. 
For xenon, the cross-section per nucleon, defined as $\sigma_{\chi p}\equiv\sigma(\chi_1+{\rm Xe}\to\chi_1+{\rm Xe})m_p^2/m_{\rm Xe}^2A^2$, where $m_p$ is the nucleon mass, is
\be
\sigma_{\chi p}=1.4\times\sin^2\xi\left(\frac{g_{\nu}^2}{10^{-2}}\right)\frac{Q^2_V}{50} \times 10^{-38}~{\rm cm}^2.
\ee
For $M_{\chi}=5$~TeV, the LUX experiment constrains $\sigma_{\chi p}<6\times 10^{-44}$~cm$^2$ at the 90\% confidence level \cite{Akerib:2013tjd}, which translates into 
\be
\sin^2\xi < 4.5 \times10^{-6}\times \left(\frac{50}{Q^2_V}\right)\left(\frac{10^{-2}}{g_{\nu}^2}\right).
\ee
The next round of direct-detection experiments, assuming that WIMPs are not detected, will start to seriously constrain $\chi_1$ as the dark matter. The estimates for the relic density and constraints from direct detection are in agreement with more general results for ``electroweakly coupled'' DM \cite{deSimone:2014pda}.

Finally, the model also predicts signals for indirect detection experiments. In regions where the density of $\chi_1$ particles is large, they can annihilate into light SM particles, right-handed neutrinos, or sterile neutrinos. In the limit discussed above, $M_{\chi}\gg M_Z,M_{\tilde{Z}}$, we expect the $\nu^c\bar{\nu^c}$ and, if kinematically accessible, $N\bar{N}$ final states to dominate. Right-handed neutrinos are virtually invisible. At high energies, the $\nu^c$ states are, for all practical purposes, massless, and interact with ordinary matter via $Z$ and $\tilde{Z}$ exchange, both cross-sections suppressed by $\sin^2\xi$ relative to that of ordinary high energy neutrino-matter scattering. The $N\bar{N}$ final states are more interesting, given that the $N$ particles will decay into high energy neutrinos and charged-leptons, as discussed in Sec.~\ref{nu_mass}. Hence, the process $\chi_1\bar{\chi}_1\to N\bar{N}$ is expected to yield high energy (energies $\lesssim M_{\chi}$) electrons, positrons, and neutrinos.


\section{Abelian Gauge Theories with Chiral Fermions}
\label{so10}


In this section we describe a general method for producing models of a Dark Sector with a fermion content that is chiral and anomaly-free (i.e. models in which there does \emph{not} exist a right-handed partner with opposite charge for every left-handed particle). This issue was also studied in detail in \cite{DobrescuSpivak}. Some of the results presented here were also discussed in \cite{DobrescuSpivak}.

Let $\msc{G}$ be a simple Lie group and $\mcl{H}$ be its Cartan subalgebra. Any $H \in \mcl{H}$ defines a $U(1)$ subgroup of $\msc{G}$. The states in an arbitrary representation $r$ transform under this $U(1)$ with charges given by the eigenvalues of $H_r$, the Cartan generator written in the representation $r$.

A sufficient condition for a gauge $U(1)$ to be anomaly free is to impose that charges and field content are in correspondence with the eigenvalues of $H_r$ for a group $\msc{G}$ that is anomaly-free. It is well known that the classical groups $SU(2)$, $SO(n)$ for $n\neq6$, $Sp(2n)$, $G_2$, $F_4$, $E_7$ and $E_8$ are all anomaly-free. For $\msc{G}$ any of these groups, a choice of representation $r$ and Cartan generator $H$ determines a $U(1)$ gauge theory that is anomaly free.

If we demand such gauge theory to be also chiral, the set of possible groups and representations is reduced considerably. All of the above groups except for those of the form $SO(4n+2)$ only have real representations. For a real representation, the matrices $H_r$ and $-H_r^*$ are similar, $H_r = U(-H_r^*)U^\dagger$ for some $U$ in $\msc{G}$. Hence, the eigenvalues of $H_r$ and $-H_r^*$ must be equal. Since $H_r$ is hermitian, its eigenvalues are real. From here we conclude that if $\lambda$ is an eigenvalue of $H_r$, then $-\lambda$ also is. Hence, real representations do not lead to chiral models.

In order to generate anomaly free chiral models one needs a group with complex representations that is anomaly free. $SO(10)$ is the smallest such group  since $SO(6)$ is anomalous. The smallest complex representation of $SO(10)$ is the spinorial $\mbf{16}$. A generator belonging to the $SO(10)$ Cartan subalgebra in this representation has the general form
\begin{align}
H(a,b,c,d,e) = \frac{1}{N}\trm{diag}\{ & a + b + c + d + e, \, -a + b + c + d - e,\, a - b + c + d - e, \, -a - b + c + d + e,  \nn \\
&  a + b - c + d - e, \, -a + b - c + d + e,\, a - b - c + d + e, \, -a - b - c + d - e, \nn \\
& a + b + c - d - e,\, -a + b + c - d + e,\, a - b + c - d + e,\, -a - b + c - d - e, \nn \\
& a + b - c - d + e, \, -a + b - c - d - e,\, a - b - c - d - e,\, -a - b - c - d + e \}   
\end{align}
where $a$, $b$, $c$, $d$ and $e$ are arbitrary real numbers and $N$ is a normalization factor. Two cases can be distinguished:
\begin{itemize}
\item At least one of $a$, $b$, $c$, $d$ or $e$ vanishes. Then, $-H$ also belongs to the Cartan subalgebra.
\item None of $a$, $b$, $c$, $d$ or $e$ vanishes. In this case, $-H$ is not a generator in this representation.
\end{itemize}
In any case, since $SO(10)$ is anomaly free, we have
\be
\tr{H} = \tr{H^3} = 0\,,
\ee
as promised. That is, as long as none of $a$, $b$, $c$, $d$ or $e$ vanishes, the elements in the diagonal above are \emph{chiral} solutions to the anomaly equations
\be
\sum_{i=1}^{16} q_i = 0\,,\quad \quad \sum_{i=1}^{16} q_i^3 = 0 , \label{anomaly-eqs}
\ee
where the $q_i$ are $U(1)$ charges with respect to a $U(1)$ gauge group.

Without too much loss of generality, we restrict the rest of our discussion to integer charges. From this requirement, either none, two, or four out of $a$, $b$, $c$, $d$ and $e$ can be half-integers, the rest must be integers. Now, notice that
\be
H(-a,b,c,d,e) = H(a,-b,c,d,e) = H(a,b,-c,d,e) = H(a,b,c,-d,e) = H(a,b,c,d,-e) = - H(a,b,c,d,e) \,.
\ee
Hence, up to an overall minus sign in all charges, it is enough to consider $a$, $b$, $c$, $d$, $e$ strictly positive. The largest $U(1)$ charge in this case is given by $a+b+c+d+e$ and notice that it appears only once. The absolute values of all other charges is necessarily smaller. Hence, for any anomaly-free model derived in this way, there can be only one state with the highest $U(1)$ charge. In particular, this state cannot be charged under any extra nonabelian gauge symmetries that commute with the $U(1)$ in question. In the SM, that state is the left-handed anti-electron $e^c$, which has hypercharge 6 in units of the smallest known hypercharge, that of the doublet quark field $Q$. Note that $e^c$ is not charged under color $SU(3)$ or electroweak $SU(2)$.

The following values
\be
a = b = c = d = \frac{1}{2}\,,\quad e = 1 \,,
\ee
yield a solution for the anomaly equations with the smallest highest charge possible, equal to 3 units in the normalization defined above, in which all charges are integers. Discarding nonchiral pairs (i.e., charges which are equal in magnitude and opposite in sign) and vanishing charges, this solution has the following particle content:
\be
\mbf{3} \times 1 \,,\quad \mbf{-2} \times 4\,,\quad \mbf{1}\times 5,
\label{model1}
\ee
where $\mbf{n}\times m$ means $m$ particles of charge $\mbf{n}$. All together there are ten chiral fields. It can be quickly checked that the anomaly conditions are satisfied.

With highest charge equal 4 we can take the $a,b,c,d,e$ combinations $\{1/2, 1/2, 1, 1, 1\}$,  $\{1/2, 1/2, 1/2, 1/2, 2\}$ or  $\{1/2, 1/2, 1/2, 3/2, 1\}$. The last of these yields a particularly simple particle content (only 7 states):
\be
\mbf{4} \times 1 \,,\quad \mbf{-3} \times 3\,,\quad \mbf{2}\times 2 \,,\quad \mbf{1}\times 1 \,.
\label{model2}
\ee
With highest charge equal to $\mbf{5}$ an even simpler particle content is obtained (6 fields):
\be
\mbf{5} \times 1 \,,\quad \mbf{-4} \times 2\,,\quad \mbf{1}\times 3 \,.
\label{model3}
\ee

Then, what is the minimum particle content possible that is chiral and allows for at least one anomaly-free $U(1)$ gauge symmetry? The following solution of Eqs.~\eqref{anomaly-eqs} with highest charge $\mbf{10}$ requires only five fields:
\be
\mbf{10}\times 1 \,,\quad \mbf{-9}\times 1\,, \quad \mbf{-7}\times 1 \,,\quad \mbf{4}\times 1\,,\quad \mbf{2}\times 1\,.
\label{model4}
\ee
On the other hand, there are no solutions to  Eqs.~\eqref{anomaly-eqs} with three fields due to Fermat's Last Theorem. It is also relatively easy to prove that for four fields only nonchiral solutions exist\footnote{We write Eq.~\eqref{anomaly-eqs} in such a way that all $q_i$ are positive. Then there are essentially two cases:
  \begin{enumerate}
  \item $q_1 = q_2 + q_3 + q_4$. Hence
    \be
    q_1^3 = (q_2 + q_3 + q_4)^3 > q_2^3 + q_3^3 + q_4^3 \nn
    \ee
    so there can be  no solutions in this case.
  \item $q_1 + q_2 = q_3 + q_4$ and $q_1^3 + q_2^3 = q_3^3 + q_4^3$. Then, one must also have $q_1q_2 = q_3q_4$. Any solution to this set of equations involves two pairs of equal charges.
    \end{enumerate}
}. Hence five chiral fields is minimal.

The model we explore in the previous sections is not minimal -- ten fields, highest charge 5 -- according to the two criteria defined above. It, however, still ``more minimal'' than the SM, which has fifteen chiral fields per generation and highest charge 6.

The model associated with Eq.~(\ref{model1}), if one assumes a Higgs-like field with charge $+1$ and a nonzero vacuum expectation value, would, in general, contain four Dirac fermions and two massless chiral fermions (linear combinations of the one $\mathbf{3}$ and five $\mathbf{1}$ states). Similarly, the model associated with Eq.~(\ref{model2}), if one assumes a Higgs-like field with charge $+1$ and a nonzero vacuum expectation value would, in general, describe three massive Dirac fermions and a massless chiral fermion, $\mathbf{1}$. This model is similar to the one described in detail in the previous sections, minus the Dirac ``sterile'' neutrino states.

It is amusing that in the case of the smallest model (five fields), associated to Eq.~(\ref{model4}) one needs more than one Higgs-like field in order to render more than one pair of chiral fermions massive. If one were to add scalar fields with charge $+1$ and $+6$ one would end up with two massive Dirac fermions and one massless state, $\mathbf{-7}$.


\section{Comments and Outlook}
\label{end}


The fermion content of the SM is chiral and before EWSB, all masses are identically zero. Only after the spontaneous breaking of the gauge symmetry does the theory allow for massive fermions and all masses are proportional to the same mass-scale. A side-effect of the chiral nature of the SM fermion content is that some fermions end up massless (i.e., the neutrinos) and some of the massive objects in the theory are, thanks to accidental global symmetries, stable (i.e., the proton). Inspired by the SM, we posit that all ``light'' fermions must be chiral as far as some gauge symmetry is concerned and therefore massless for as long as the symmetry is manifest. Masses for all such fermions should come as the result of a Higgs mechanism that spontaneously breaks the gauge symmetry at some energy scale.

The scenario described above invites two main questions. First, can one construct models within this paradigm generically and systematically? How? Second, are these models useful to address outstanding puzzles in fundamental physics, including the dark matter puzzle and the origin of nonzero neutrino masses? In the bulk of this paper, we dealt with the second question. We fleshed out a model of a dark sector (DS), consisting of a new, spontaneously broken $U(1)_{\nu}$, along with chiral fermions charged under $U(1)_{\nu}$. This model indeed leads to a candidate DM particle in the form of a new WIMP-like massive stable state. A relic abundance for that state appears as a result of a standard thermal freeze-out mechanism occuring in the DS. We showed that the $U(1)_{\nu}$ breaking scale can be chosen such that the relic density of the WIMP-like object is consistent with the evidence for dark matter in the universe.   

The sample-model also contains massless chiral fermions, which as far as the SM gauge group is concerned, look like right-handed neutrinos. Nonzero neutrino masses require the introduction of a new, potentially very heavy, mediator mass scale that connects these states with the massless left-handed neutrinos of the SM. We discussed in detail an ultraviolet complete scenario that preserves lepton number but nonetheless leads to parametrically small Dirac neutrino masses. $U(1)_{\nu}$ anomaly constraints also require the existence of at least four mostly sterile Dirac neutrinos -- two sterile neutrinos are required per generation of new fermions, and the neutrino data require at least two of these generations -- that mix with the SM neutrinos. The mixing between active and sterile neutrinos is related, qualitatively, to the observed nonzero active neutrino masses.

Some extra features of the model might be considered appealing. The scalar sector of the DS is minimal in the sense that one field is enough to give mass to all the DS particles except for the RH neutrino. Moreover, the fermion content is also minimal in the sense that only two states are stable on cosmological scales, the DM and the RH neutrino. Thus, the only stable particle that the model adds to the SM ones is the DM fermion. All in all, this model is rich enough to provide DM, neutrino masses and some interesting phenomenology, and yet it is simple enough not to modify the SM dramatically.

For the sake of concreteness, we brushed aside several possibilities that may be worthy of further exploration. For this model, we chose a mediator sector that consisted of a new vector-like fermion charged under both the SM and the $U(1)_{\nu}$ gauge symmetries. A different mediator sector would lead to other sources for nonzero masses for both the SM and the DS ``neutrinos,'' including some that would render all of these particles massive Majorana fermions. Different possibilities can be explored qualitatively by appreciating that there are two dimension-five operators other than Eq.~(\ref{eq:dim-5}) one can construct, given the particle content and the gauge symmetries discussed in detail here. Explicitly, one may have $(LH)(LH)$, the so-called Weinberg operator, and $(\mathbf{1}_+\phi)(\mathbf{1}_+\phi)$. The former leads to nonzero Majorana masses for the active neutrinos, proportional to $v^2/M_{\slashed{L}}$ while the latter leads to ``right-handed neutrino'' masses proportional to $v_{\phi}^2/M_{\slashed{L}}$, where $M_{\slashed{L}}$ is the generalized lepton-number-breaking scale. This possibility was considered recently in \cite{Roland:2014vba}. The scale $M_{\slashed{L}}$ is, in general, not related to $\Lambda$, the mass of the mediator field $X$. If $M_{\slashed{L}}\gg\Lambda$, the results discussed here remain valid. This would happen, for example, for $M_{\slashed{L}}\sim 10^{16}$~GeV, the grand-unification scale. In this case, both left-handed and right-handed neutrinos receive nonzero Majorana masses, but these are much smaller than the Dirac masses and the neutrinos would be pseudoDirac fermions that, for most experimental purposes, ``look like'' Dirac neutrinos.

More speculatively, while we extended the SM gauge symmetry with a new $U(1)_{\nu}$, more complicated DS gauge symmetries are also a possibility worthy of pursuit. In particular, it is easy to check that in the model discussed in detail in this paper, one can add a new, nonanomalous\footnote{As far as the $SU(2)_{\nu}$ is concerned, the fermions are vector-like.} $SU(2)_{\nu}$ gauge symmetry, under which $(\mathbf{1}^1_+,\mathbf{1}^2_+)$,  $(\mathbf{2}^1_-,\mathbf{2}^2_-)$, $(\mathbf{3}^1_-,\mathbf{3}^2_-)$, $(\mathbf{4}^1_+,\mathbf{4}^2_+)$ all transform as doublets. This $SU(2)_{\nu}$ is not spontaneously broken by $v_{\phi}$, since $\phi$ is an $SU(2)_{\nu}$ singlet. There would be, nonetheless, four massive Dirac fermions per DS generation, pairwise degenerate due to the unbroken $SU(2)_{\nu}$ symmetry. At low energies and for a small enough number of dark matter generations, the $SU(2)_{\nu}$ interaction would confine and those states charged under it would manifest themselves as dark ``mesons'' or ``baryons''. In particular, the lightest, stable such state could be the dark matter\footnote{The case where the $SU(2)_{\nu}$ is infrared free is probably ruled out by the fact that long-range dark matter self-interactions are excluded by observations.}. Note that this $SU(2)_{\nu}$ cannot be naively identified as the SM $SU(2)_L$ since, if this were the case, all members of DS doublets would acquire electric charge $\pm1/2$ after EWSB. 

It is interesting to enquire whether the model proposed here is natural, at least as far as it is defined in, for example, \cite{Farina:2013mla,deGouvea:2014xba}. In more detail, naturalness in this context translates into requiring that finite loop-corrections to the masses of $H$ and $\phi$ do not supersede the masses themselves. The presence of the messenger field $X$, for example, can lead to very large corrections to the masses of both scalars. At the one loop-level, naturalness translates into, roughly,
\begin{equation}
\frac{\kappa^2_L}{8\pi^2}\Lambda^2,~ \frac{\kappa^2_0}{8\pi^2}\Lambda^2 \lesssim (100~{\rm GeV})^2.
\end{equation}
According to this criterium, keeping in mind that the active neutrino masses, of order $10^{-1}$~eV, are proportional to $\kappa_0\kappa_L/\Lambda$, the theory is natural -- in the sense that the $\Lambda$ mass scale does not destabilize the electroweak and $U(1)_{\nu}$ breaking scales --  as long as $\Lambda\lesssim 10^7$~GeV (assuming $\kappa_L\sim \kappa_0$ and $v\sim v_\phi\sim 10^2$~GeV).

Independent from the mediator sector, the presence of two mass-scales, $v$ and $v_\phi$ is ``unnatural'' \cite{Farina:2013mla,deGouvea:2014xba}  unless $v\sim v_\phi$ or if the coupling $\lambda_{\phi H}$ between the two sector is very small. Naturalness from the neutrino sector naively requires, as discussed above, $\kappa_{i,L}\lesssim 10^{-4}$, and one can hence argue that tiny $\lambda_{\phi H}$ values are, in some sense, expected (see discussion in Sec.~\ref{new_u1} and Eq.~(\ref{lambdaphiH})). This implies that even if one takes naturalness considerations into account, $v$ can be significantly smaller (or larger) than $v_\phi$. 

Many other potential consequences of the model were not discussed here. Since the neutrinos are Dirac fermions, standard leptogenesis does not work. It is, therefore, reasonable to ask whether the scenario discussed here can accommodate a baryogenesis mechanism. For example, would Dirac leptogenesis~\cite{Dick:1999je} work? We also did not consider the possibility that the $U(1)_{\nu}$ gauge boson is light ($M_{\tilde{Z}}\lesssim 1$~GeV) or very light ($M_{\tilde{Z}}\ll 1$~MeV), nor did we explore the consequences of postulating that the dark matter is light. We also did not consider the possibility that the sterile neutrino states -- whether they be hot, cold, or warm --  could make up most or part of the dark matter.

Finally, we return to the first question posed above: how to construct such models generically and systematically? In Sec.~\ref{so10}, we discussed a simple yet powerful way of generating an infinite number of chiral, anomaly-free $U(1)$ gauge theories, of which our model is only one example. Each model found by this method will have its own idiosyncrasies and many should allow one to address the current puzzles in particle physics. It also seems more-or-less straightforward to use the same $SO(10)$-inspired mechanism in order to identify anomaly-free theories with chiral fermions that are invariant under larger gauge symmetries, like the SM.


\section*{Acknowledgments}

We thank Janet Conrad for asking the questions that set us thinking about new ``neutrino'' gauge interactions. This work is sponsored in part by the US Department of Energy Contract DE-FG02-91ER40684.

\end{document}